\documentstyle[12pt,psfig]{article}
\begin{document}

\begin{titlepage}
\begin{center}

{\Large\bf{The improved nuclear parton distributions}}
\\[5.0ex]
{\Large\it{A. L. Ayala Filho $^{*}$\footnotetext{$^{*}$E-mail:ayala@ufpel.tche.br}}}
 {\it and}
{ \Large \it{ V. P.  Gon\c{c}alves
$^{**}$\footnotetext{$^{**}$E-mail:barros@ufpel.tche.br} }}
\\[1.5ex]
 {\it  Instituto de F\'{\i}sica e Matem\'atica, Univ. Federal
de Pelotas}\\ {\it Caixa Postal 354, 96010-090 Pelotas, RS, BRAZIL}\\[5.0ex]
\end{center}

{\large \bf Abstract:} In this paper  we propose an improvement
 of the EKS nuclear parton
distributions for the small $x$ region of high energy processes, where the
perturbative high parton density  effects cannot be disregarded. We analyze
the behavior of the ratios $xG_A/xG_N$ and $F_2^A/F_2^D$ and  verify that at small $x$
they are strongly modified when compared to the EKS predictions. The implications of our results for the heavy ion
collisions in RHIC and LHC are discussed.

\vspace{1.5cm}



\end{titlepage}

\section{Introduction}
\label{intro}

The physics of high-density QCD has become an increasingly active subject of
research, both from experimental and theoretical points of view. In
particular, the collider facilities such as the BNL Relativistic Heavy Ion
Collider (RHIC), and CERN Large Hadron Collider (LHC) will be able to probe
new regimes of dense quark matter at very small Bjorken $x$ or/and at large $%
A$, with rather different dynamical properties. Basically, new phenomena
associated with an ultradense environment that may be created in the central
collision region of these reactions are expected \cite{harris}.

From the analysis of nucleus-nucleus collisions for RHIC energies and
beyond, we have that the perturbative QCD processes should determine the
initial conditions, with most of the entropy and transverse energy
presumably produced already during very early times (within the first 2 $fm$
after the nuclear contact) by frequent, mostly inelastic, semihard gluonic
collisions involving typical momentum transfers of only a few $GeV$ \cite
{eskm}. In particular, at early times, $\tau \approx 1/p_T\le 1/p_0\approx
0.1\,fm$ for $p_0\approx 2\,GeV$, semihard production of minijets will set
the stage for further evolution of the system. The calculation of this
process is based on the jet cross section for $p_T>p_0$, with the parton
densities evaluated at scale $p_T$, with $x$ values at central rapidities as
low as $x\approx {\cal {O}}(10^{-4})$ in $Pb+Pb$ collisions at LHC ($%
5.5\,TeV/nucleon)$. At the lower RHIC energies, $x\approx {\cal {O}}%
(10^{-2}) $ at central rapidities, and at higher rapidities the $x$ values
probed can be even smaller. Thus the small $x$ behavior of the parton
densities strongly influences the initial conditions of the minijet system.

While the deep inelastic scattering data from HERA continues to refine the
parton densities at small $x$, uncertainties in the distributions still
exist, mainly associated to the high parton density effects present in this
kinematical region. Such effects would be present in nucleus-nucleus ($AA$)
collisions at collider energies, modifying the perturbative QCD predictions
for global observables, such as particle multiplicities and transverse
energy production, as well as minijet production, heavy quarks, their bound
states, and dilepton production \cite{rep}. Consequently, the high density
effects are one of the major theoretical issues in modeling the QCD
processes in nuclear collisions.

In this paper we  analyze the high density effects in the behavior of
the nuclear gluon distribution $xG_A$ and nuclear structure function $F_2^A$
at small $x$ and a large perturbative scale $Q^2$. Our study is motivated by
the perspective that in a near future an experimental investigation of these
effects at small $x$ and $Q^2>1\,GeV^2$ using $eA$ scattering could occur in
RHIC, as well as at DESY Hadron Electron Ring Accelerator (HERA).
Furthermore, our goal is to improve the description proposed by Eskola,
Kolhinen and Salgado (EKS) taking into account the high parton densities effects
 in the parton evolution at small $x$.

In recent years several experiments have been dedicated to high precision
measurements of deep inelastic lepton scattering (DIS) off nuclei.
Experiments at CERN and Fermilab focus especially on the region of small
values of the Bjorken variable $x = Q^2/2M\nu$, where $Q^2=-q^2$ is the
squared four-momentum transfer, $\nu$ the energy transfer and $M$ the
nucleon mass. The data \cite{arneodo,e665}, taken over a wide kinematic
range, have shown that the proton and neutron structure functions are
modified by a nuclear environment. The modifications depends on the parton
momentum fraction: for momentum fractions $x < 0.1$ and $0.3 < x < 0.7$, a
depletion is observed in the nuclear structure functions. The low $x$
(shadowing region) and the larger $x$ (EMC region) are bridged by an
enhancement known as antishadowing for $0.1 < x < 0.3$. We refer to the
entire phenomena as {\it the nuclear shadowing effect}.

The theoretical understanding of $F_2^A$ in the full kinematic region has
progressed in recent years, with several models which describe the
experimental data with quite success \cite{reviews}. Here we will
restrict ourselves to the descriptions which use the DGLAP evolution
equations \cite{dglap} to describe the behavior of the nuclear parton
distributions. Recently, Eskola, Kolhinen and Salgado \cite{eks}, following
Ref. \cite{eskola}, have shown that the experimental results \cite
{arneodo} presenting nuclear shadowing effects can be described using the
DGLAP evolution equations \cite{dglap}
with adjusted initial parton distributions. The basic idea of this framework
is the same as in the global analyzes of parton distributions in the free
proton: they determine the nuclear parton densities at a wide range of $x$
and $Q^2\geq Q_0^2=2.25$ GeV$^2$ through their perturbative DGLAP evolution
by using the available experimental data from $lA$ DIS and Drell-Yan (DY)
measurements in $pA$ collisions as constraint. EKS have expressed the
results in terms of the nuclear ratios $R_f^A(x,Q^2)$ for each parton flavor 
$f$ in a nucleus with $A$ nucleons ($A>2$), at $10^{-6}\leq x\leq 1$ and $%
2.25\,GeV^2\leq Q^2\leq 10^4\,GeV^2$. The results of EKS
seems to show that, in the kinematic region of the present data, the high density
dynamical effects are small enough to be described by the DGLAP evolution
equation with a suitable set of nonperturbative initial conditions. 
The main shortcoming of use the EKS parameterizations in the
calculations of $eA$, $pA$ or $AA$ processes is associated to the small $x$
predictions, where the solution of the DGLAP equations reduces to the well
known limit of the double logarithm approximation (DLA). This limit is
characterized by a strong growth of the gluon distribution, with a similar
behavior for the $F_2$ structure function, which implies a high parton density in
this kinematic region. However, when the density of gluons and quarks
becomes very high the physical processes of interaction and recombination of
partons, not considered in the DGLAP equations, become important in the
parton cascade and these effects should be expressed in a new evolution
equation. Therefore, the EKS description should  be improved to include
the high parton density effects when smaller values of $x$ are considered. This is
the case, for instance, in the calculation of the minijet cross section at LHC.

At this moment, there are many approaches in the literature that propose
distinct evolution equations for the description of the gluon distribution
in high density limit \cite{100,ayala1} \cite{jamal,kov}. In general these
evolution equations resum powers of the function $\kappa (x,Q^2)\equiv \frac{%
3\pi ^2\alpha _sA}{2Q^2}\frac{xg(x,Q^2)}{\pi R_A^2}$, which represents the
probability of gluon-gluon interaction inside the parton cascade. Moreover,
these equations match (a) the DLA limit of the DGLAP evolution equation in
the limit of low parton densities $(\kappa \rightarrow 0)$; (b) the GLR
equation and the Glauber-Mueller formula as first terms of the high density
effects. The main differences between these approaches are present in the limit
of very large densities, where all powers of $\kappa $ should be resumed.
Although the complete demonstration of the equivalence between these
formulations in the region of large $\kappa $ is still an open question,
some steps in this direction were given recently \cite{npbvic,kovner}. Here
we will consider the Glauber-Mueller approach for the high density effects,
which is a common limit of the current high density approaches in the
kinematic region which we are interested. Thus we intend to obtain no model
dependent predictions.

The outline of this paper is the following. In next section we present 
 a brief review of  the   Glauber-Mueller approach for the
nuclear structure function (For details see \cite{ayala1}). In Sec. \ref{ekspar} we analyze 
the EKS parameterization and present a procedure to improve this paramerization for the small $x$ region,
where the high density effects cannot be disregarded. Moreover, we present our results for 
the    ratios $xG_A/xG_N$ and $F_2^A/F_2^D$ and  verify that at small $x$
they are strongly modified in comparison with the EKS predictions. Finally, in Sec. \ref{conclusions} we 
discuss the implications of our results for the heavy ion collisions in RHIC and LHC and present our 
conclusions.   

\section{The High Density Effects in DIS}

The deep
inelastic scattering off a nucleus is usually interpreted in a frame where
the nucleus is going very fast. In this case the nuclear shadowing is a
result of an overlap in the longitudinal direction of the parton clouds
originated from different bound nucleons \cite{100}. Thus low $x$ partons
from different nucleons overlap spatially creating much larger parton
densities than in the free nucleon case. This leads to a large amplification
of the nonlinear effects expected in QCD at small $x$. In the target rest
frame, the electron-nucleus scattering can be visualized in terms of the
propagation of a small $q\overline{q}$ pair in high density gluon fields
through much larger distances than it is possible with free nucleons. In
terms of Fock states we then view the $eA$ scattering as follows \cite
{gribov}: the electron emits a photon ($|e>\rightarrow |e\gamma >$) with $%
E_\gamma =\nu $ and $p_{t\,\gamma }^2\approx Q^2$, after the photon splits
into a $q\overline{q}$ ($|e\gamma >\rightarrow |eq\overline{q}>$) and
typically travels a distance $l_c\approx 1/m_Nx$, referred as the 'coherence
length', before interacting in the nucleus. For small $x$ (large $s$, where $%
\sqrt{s}$ is the $\gamma ^{*}A$ center-of-mass energy), the photon converts to a
quark pair at a large distance before it interacts to the target.
Consequently, the space-time picture of the DIS in the target rest frame can
be viewed as the decay of the virtual photon at high energy (small $x$) into
a quark-antiquark pair, which subsequently interacts with the target. In the
small $x$ region, where $x\ll \frac 1{2mR}$, the $q\overline{q}$ pair
crosses the target with fixed transverse distance $r_t$ between the quarks.
Following Gribov \cite{gribov}, we may write a double dispersion relation
for the forward $\gamma ^{*}A$ elastic amplitude ${\cal A}$, related to the
total cross section by the optical theorem [$Im\,{\cal A}=s\sigma (s,Q^2)$],
and obtain for fixed $s$ 
\begin{eqnarray}
\sigma (s,Q^2)=\sum_q\int \frac{dM^2}{M^2+Q^2}\frac{dM^{\prime 2}}{M^{\prime
2}+Q^2}\rho (s,M^2,M^{\prime 2})\frac 1s\,Im\,{\cal A}_{q\overline{q}%
+A}(s,M^2,M^{\prime 2})\,\,,  \label{sigdis}
\end{eqnarray}
where $M$ and $M^{\prime }$ are the invariant masses of the incoming and
outgoing $q\overline{q}$ pair. If we assume that forward $q\overline{q}+A$
scattering does not change the momentum of the quarks then ${\cal{A}}_{q\overline{q}%
+A}$ is proportional to $\delta (M^2-M^{\prime 2})$, and (\ref{sigdis})
becomes 
\begin{eqnarray}
\sigma (s,Q^2)=\sum_q\int \frac{dM^2}{(M^2+Q^2)^2}\rho (s,M^2)\sigma _{q%
\overline{q}+A}(s,M^2)\,\,,  \label{sigdis2}
\end{eqnarray}
where the spectral function $\rho (s,M^2)$ is the density of $q\overline{q}$
states, which may be expressed in terms of the $\gamma ^{*}\rightarrow q%
\overline{q}$ matrix element \cite{ryskin}. Using that $%
M^2=(k_t^2+m_q^2)/[z(1-z)]$, where $k_t$ and $z$ are the transverse and
longitudinal momentum components of the quark with mass $m_q$, we can
express the integral over the mass $M$ of the $q\overline{q}$ in terms of a
two-dimensional integral over $z$ and $k_t$. Instead of $k_t$ is useful to  work
with the transverse coordinate $r_t$ (impact parameter representation),
which is the variable Fourier conjugate to $k_t$, resulting \cite{nik} 
\begin{eqnarray}
F_2^A(x,Q^2) &=&\frac{Q^2}{4\pi \alpha _{em}}\sigma (s,Q^2)  \nonumber \\
&=&\frac{Q^2}{4\pi \alpha _{em}}\int dz\int \frac{d^2r_t}\pi |\Psi
(z,r_t)|^2\,\sigma _{q\overline{q}+A}(z,r_t)\,\,,  \label{f2target}
\end{eqnarray}
where 
\begin{eqnarray}
|\Psi (z,r_t)|^2=\frac{6\alpha _{em}}{(2\pi )^2}\sum_f^{n_f}e_f^2%
\{[z^2+(1-z)^2]\epsilon ^2\,K_1(\epsilon r_t)^2+m_i^2\,K_0(\epsilon
r_t)^2\}\,\,.  \label{wave}
\end{eqnarray}
The photon wave function $\Psi (z,r_t)$ is simply the Fourier transform of
the matrix element for the transition $\gamma ^{*}\rightarrow q\overline{q}$%
. Moreover, $\alpha _{em}$ is the electromagnetic coupling constant, $%
\epsilon ^2=z(1-z)Q^2+m_i^2$, $m_i$ is the quark mass, $n_f$ is the number
of active flavors, $e_f^2$ is the square of the parton charge (in units of $%
e $), $K_{0,1}$ are the modified Bessel functions and $z$ is the fraction of
the photon's light-cone momentum carried by one of the quarks of the pair.
In the leading log$(1/x)$ approximation we can neglect the change of $z$
during the interaction and describe the cross section $\sigma ^{q\overline{q}%
}(z,4/r_t^2)$ as a function of the variable $x$. To estimate the high
density effects we consider the Glauber multiple scattering theory \cite
{chou}, which was derived in QCD \cite{muegla}. In this framework the nuclear
collision is analyzed as a succession of independent collisions of the probe
with individual nucleons within the nucleus, which implies that 
\begin{eqnarray}
F_2^A(x,Q^2)=\frac{Q^2}{4\pi \alpha _{em}}\int dz\int \frac{d^2\vec{r}_t}\pi
|\Psi (z,\vec{r}_t)|^2\,\int \frac{d^2\vec{b}_t}\pi \,2\,[1-e^{-\sigma ^{q%
\overline{q}+N}(z,\vec{r}_t)S(\vec{b}_t)}]\,\,,  \label{siga}
\end{eqnarray}
where $\vec{b}_t$ is the impact parameter, $S(\vec{b}_t)$ is the profile
function and $\sigma ^{q\overline{q}+N}$ is the dipole cross section off the
nucleons inside the nucleus, which is proportional to the pair separation
squared $r_t^2$ and the nucleon gluon distribution $xg(x,1/r_t^2)$. The
expression (\ref{siga}) represents  the Glauber-Mueller formula for the
nuclear structure function (see \cite{ayala1} for details). The use of a
Gaussian parameterization for the nucleon profile function $S(b_t)=\frac
1{\pi R_A^2}e^{-\frac{b^2}{R_A^2}}$, where $R_A$ is the mean nuclear
radius, simplifies the calculations. We obtain that the $F_2^A$ structure
function can be written as \cite{ayala1} 
\begin{eqnarray}
F_2^A(x,Q^2)=\frac{R_A^2}{2\pi ^2}\sum_{u,d,s}\epsilon _i^2\int_{\frac
1{Q^2}}^{\frac 1{Q_0^2}}\frac{d^2r_t}{\pi r_t^4}\{C+ln(\kappa
_q(x,r_t^2))+E_1(\kappa _q(x,r_t^2))\}\,\,,  \label{diseik2}
\end{eqnarray}
where $\kappa _q=(2\alpha _sA/3R^2)\,\pi \,r_t^2\,xG_N(x,\frac 1{r_t^2})$, $%
C $ is the Euler constant, $E_1$ is the exponential integral function and $A$ the
number of nucleons in a nucleus. This equation allows to estimate the high
density corrections to the structure function in the DLA limit. Expanding
the equation (\ref{diseik2}) for small $\kappa _q$, the first term (Born
term) will correspond to the usual DGLAP equation in the small $x$ region.

The Glauber - Mueller formula has been used in a comprehensive phenomenological analyzes 
of the behavior of distinct observables in $ep$ and $eA$ processes. The results of these studies
agree with the current  $ep$ HERA data \cite{ayaepj,ayavic} and  allow us to
make some predictions which will be investigated  in the near future in $eA$
colliders \cite{vicprc,vicgay,vic}. The main shortcoming in the studies of
 $eA$ processes is that the large $x$ effects (the antishadowing and the EMC effect) were disregarded, which implies predictions only for
the asymptotic behavior (large $s$/small $x$) of the observables. In next section we propose a procedure to obtain more realistic predictions in the full kinematic region.

\section{The improved EKS parton distributions}
\label{ekspar}

As discussed in  Sec. \ref{intro} the EKS parameterization, although describes the
current experimental fixed target data quite well, is not a good approximation for the
small $x$ region in the perturbative regime, where the DGLAP equations
predicts a strong growth of the parton distributions (the DLA\ limit). In
this limit the nuclear gluon distribution is given by $xG_A\propto \exp 
\sqrt{\ln (1/x)\,\ln (Q^2/Q_0^2)}$, which is almost independent of the
initial nonperturbative input, i.e. of the adjusted parameters obtained in
the EKS parameterization. To improve the EKS description we propose
the substitution of DLA limit of the DGLAP evolution, present
in the EKS framework, by the
Glauber-Mueller evolution. This procedure implies  to  calculate
the nuclear structure
function using the following expression
\begin{eqnarray}
F_2^A(x,Q^2) & = & F_2^A(x,Q^2)[\mbox{EKS}]\, - \, (1/A) \, F_2^A(x,Q^2)[\mbox{DLA}%
]\, \nonumber \\ & + & \,(1/A)\,F_2^A(x,Q^2)[\mbox{eq. (\ref{diseik2})}],  \label{f2abc}
\end{eqnarray}
with $F_2^A(x,Q^2)[\mbox{EKS}]\,\,=R_{F_2}^A\times \,F_2^N(x,Q^2)$, 
where $R_{F_2}^A$ is obtained in terms of a combination of nuclear
parton ratios $R_f^A(x,Q^2)$ \cite{eskola}. 
The nucleon structure function  is given by
$F_2^N(x,Q^2) = \sum_{u,d,s} \epsilon_q^2 \,[xq(x,Q^2) + x\overline{q}(x,Q^2)] + F_2^c(x,Q^2)$, 
where the  charm component of the nucleon structure function is calculated
considering the charm production via boson-gluon fusion mechanism \cite{grv95} and the 
 nucleon parton distributions  are given by the  GRV parameterization \cite{grv95}.
In this  work we assume $m_c = 1.5\,GeV$.
Moreover, 
$F_2^A(x,Q^2)[\mbox{DLA}]$ is
calculated using the parton distribution from DLA limit of DGLAP evolution. For
practical purpose, this term is given by the Born term of Eq. (\ref{diseik2}).

The above   procedure implies the inclusion of: (a) the full DGLAP evolution
equation in all kinematic region; (b) the nonperturbative nuclear
corrections in the nuclear parton distributions, that describes the experimental
fixed target data,  and (c) the high density
effects present in the parton evolution at small $x$ in the perturbative regime.
Therefore, with the Eq. (\ref{f2abc}) we are able to describe the parton evolution in all
kinematic region.

A similar description can be used to estimate the nuclear gluon distribution $%
xG_A$. In this case we take
\begin{eqnarray}
xG_A(x,Q^2) & = & xG_A(x,Q^2)[\mbox{EKS}]\, - \,(1/A)\,xG_A(x,Q^2)[\mbox{DLA}%
]+ \nonumber \\ & + & \, (1/A)\,xG_A(x,Q^2)[\mbox{GM}]   \label{xgab}
\end{eqnarray}
where $xG_A(x,Q^2)\,[\mbox{EKS}] \,\,\,=\,\,\, R_G^A \,\,\times \,xG_N(x,Q^2)\,$
is the EKS prediction,  $xG_A(x,Q^2)\,[\mbox{DLA}]$ is the DGLAP (DLA)
prediction for the nuclear gluon distribution and GM represents the
Glauber-Mueller nuclear gluon distribution given by \cite{ayala1}
\begin{eqnarray}
xG_A(x,Q^2)[\mbox{GM}]=\frac{2R_A^2}{\pi ^2}\int_x^1\frac{dx^{\prime }}{%
x^{\prime }}\int_{\frac 1{Q^2}}^{\frac 1{Q_0^2}}\frac{d^2r_t}{\pi r_t^4}%
\{C+ln(\kappa _G(x^{\prime },r_t^2))+E_1(\kappa _G(x^{\prime
},r_t^2))\}\,\,\,\,,  \label{master}
\end{eqnarray}
where $\kappa _G=(9/4)\kappa _q.$
The main difference between the high density effects in the quark and gluon
densities stems from the much larger cross section $\sigma
_N^{gg}=(9/4)\sigma _N^{q\overline{q}}$, {\it i. e.} $\kappa _G=(9/4)\kappa
_q$, which in turn leads to a much larger gluon shadowing.       As before,
 the DLA limit is obtained from the Born term of the Glauber-Mueller evolution
 [Eq. (\ref{master})].

Using the above procedure we can calculate the ratios  $R^{F_2}= F_2^A/F_2^D$ and
$R_G = xG_A/xG_N$, where $F_2^D$ is the deuterium structure function. 
In order to include only perturbative contributions to the estimative of the
high density effects, we take the  initial scale at $Q_0^2 = 1 \,
GeV^2$  in the Glauber-Mueller expressions [Eqs. (\ref{diseik2}) and (\ref{master})]
and calculate the ratios for $A  = 208$.

In Fig. \ref{fig1} we present our predictions for the behavior of the ratio  $R^{F_2}= F_2^A/F_2^D$
 (denoted as EKS MOD) for two values of virtualities ($Q^2 = 2.25$ and $15 \,GeV^2$).
 For comparison the EKS predictions are also shown.
We verify that while for the region of large values of $x$ ($\ge 10^{-2}$) our predictions are almost identical to the EKS results, at small $x$ the differences are large. A comment related to experimental data is in order. 
As $x \approx Q^2/s$, where $s$ is the squared CM energy, the   data
in the region of small $x$ values are for small values of $Q^2$ ($ \le 1\,GeV^2$),
where the use of the perturbative QCD cannot be justified and the
shadowing corrections are dominated by soft contributions. Therefore, our
perturbative predictions cannot be compared with the current experimental fixed
target data.
In the perturbative regime $Q^2 \ge 1.0\, GeV^2$, where the data are
associated with $x$ values greater than $10^{-2}$,  our results are almost
identical to the EKS predictions. Thus, the fixed target data  does not
allow us to discriminate between the predictions.

The results shown in Fig. \ref{fig1} predicts a sizeable modification in the
quark distribution at low $x$ ($x \le 10^{-3}$). Thus, we expect that the $lA$
cross section will present a strong reduction when compared to the $lp$ in this
kinematic region. Also the Drell-Yan production in heavy ion collisions should
be strongly modified when compared to the $pp$ one at LHC energy ($x \approx 10^{-4}$).
At RHIC kinematic region, the high density effects does not seem to be strong
enough to modify the Drell-Yan cross section in respect to the EKS prediction.
Nevertheless, this subject deserves a further more detailed study.

The high density effects are important already for the initial scale of
the EKS parameterizations ($Q^2 = 2.25 \, GeV^2$) and increases with the
virtuality. The remarkable feature in the improved parton distributions is
the nonsaturation of the ratio at small values $x$. In the EKS
parameterization the saturation
is included in the initial condition and this general behavior is not
modified by the evolution. This is  a consequence of the DGLAP
evolution equations, which reduces to the DLA limit at small $x$ in the
nuclear and nucleon case, keeping the ratio approximately constant.
When the high density effects are considered the ratio is not constant
since these effects increases at small values of $x$ and are larger in
the nuclear case. Therefore, the nonsaturation of the ratio in the perturbative regime is a signature of the high density effects in the nuclear processes.

In Fig. \ref{fig2}  we present our predictions for the behavior of the ratio
  $R_G(x,Q^2)=xG_A/xG_N$
 (denoted as EKS MOD) for two values of virtualities
 ($Q^2 = 2.25$ and $15 \,GeV^2$). For comparison the EKS predictions are
 also shown. We verify that while for the region of large values of $x$
 ($\ge 10^{-2}$) our predictions are almost identical to the EKS results,
 at small $x$ the differences are large.   We can see from the both figures that the
 EKS assumption that  $R_G = R^{F_2}$ at small values of  $x$ is strongly
 modified by the high density effects and its $Q^2$ evolution, since
  $R_G \ll R^{F_2}$ for $Q^2 = 15 \, GeV^2$. Therefore, we predict
  a large modification in the quarkonium production at LHC, as well as in
  its  bound states (e.g. $J/\Psi ,\, \Psi^{\prime} ,\, \Upsilon ,\, ...$).
 Furthermore, as the high density effects are important already to
 $Q^2 \approx 2.25 \, GeV^2$, also the minijet production will be modified
 at LHC.    At RHIC kinematical region the high density effects does not
 significantly modify the gluon distribution
 and we expect a similar prediction of the quarkonium production from both models.
 Anyway, the high density effects should be considered in detail before to use
 the quarkonium production, as well as its bound states, as probes of the
 deconfined state of matter. Also this subject deserves a further detailed study.

It is important to note  that the EKS description use as a constraint the momentum sum rule. 
We verify that the improved distributions violates this sum rule at must in 5 $\%$ for the 
gluon distribution, which is small when compared with the experimental uncertainty on  this 
distribution in the antishadowing region. Moreover, the violation is small since the main contribution to 
the sum rule comes from the large $x$ region, where the high density effects are negligible.

\section{Conclusions}
\label{conclusions}

 In this paper we have proposed an improvement of the EKS nuclear
 parton distributions in the small $x$ region by the inclusion of the perturbative high density effects. 
We predict the nonsaturation of the ratios $R^{F_2}$ and $R_G$ when  these  effects are considered and that $R^{F_2} \ll R_G$ in the small $x$ limit.  Such results could be tested in the future $eA$ colliders. Furthermore, 
our results demonstrate that the  high density  effects are very important mainly at the LHC kinematic region, where strong modifications in the Drell- Yan and quarkonium production are expected.  As  the small $x$ behavior of the parton
densities strongly determines the initial conditions of the minijet system in  nucleus-nucleus collisions, our results show that the high density effects cannot be disregarded in the calculations of the observables and signatures of a Quark-Gluon Plasma.

\section*{Acknowledgments}

This work was partially financed by FAPERGS and CNPq, BRAZIL.

\newpage

\section*{Figure Captions}

\begin{figure}[tbp]
\centerline{\psfig{file=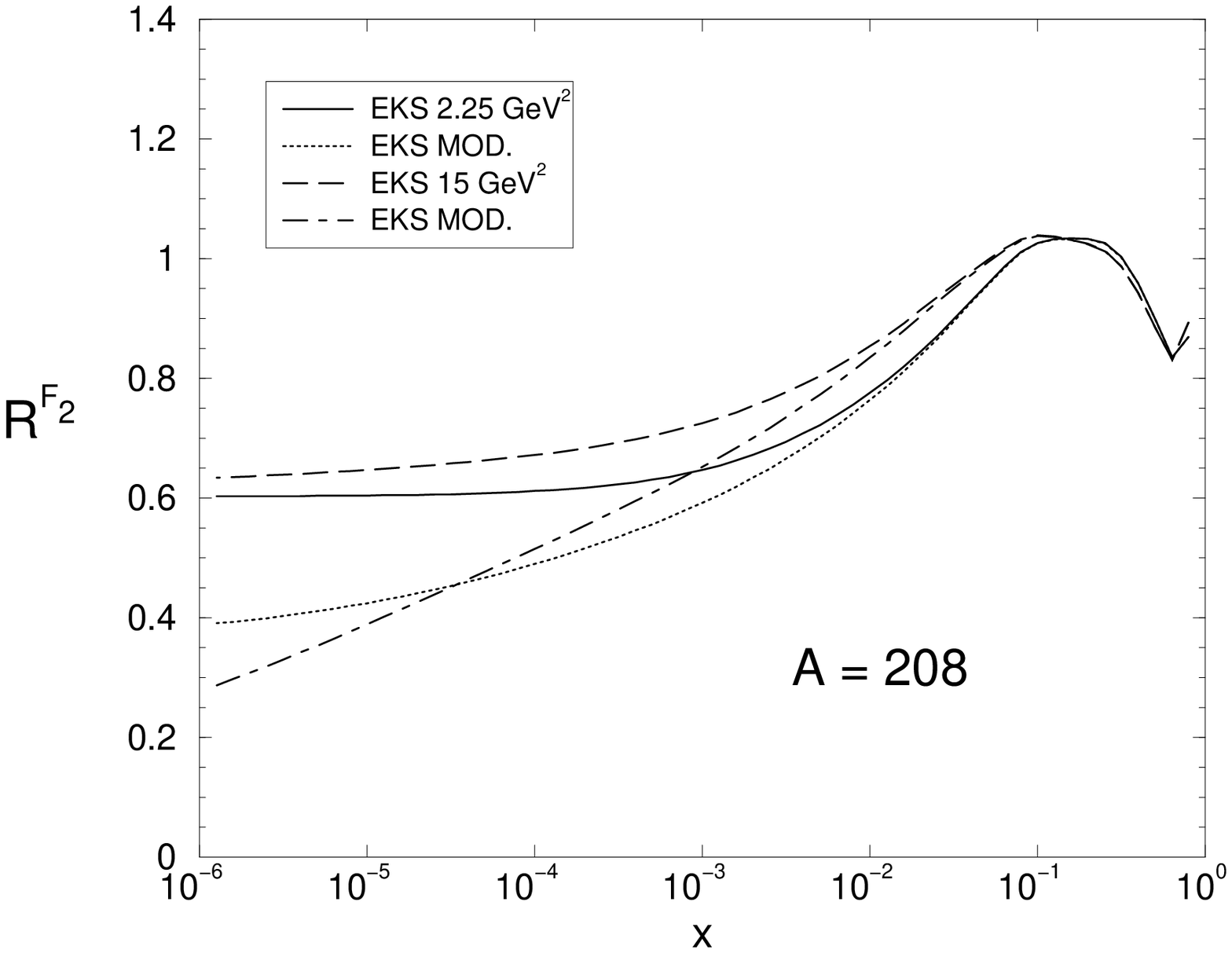,width=150mm}}
\caption{ }
\label{fig1}
\end{figure}

\vspace{1.0cm} Fig. \ref{fig1}: Comparison between the predictions of the
original EKS and EKS modified (denoted EKS MOD. in  plot) for the ratio $%
R^{F_2}(x,Q^2)=F_2^A/F_2^D$ as function of the variable $x$ at different
values of $Q^2$. See text.

\begin{figure}[tbp]
\centerline{\psfig{file=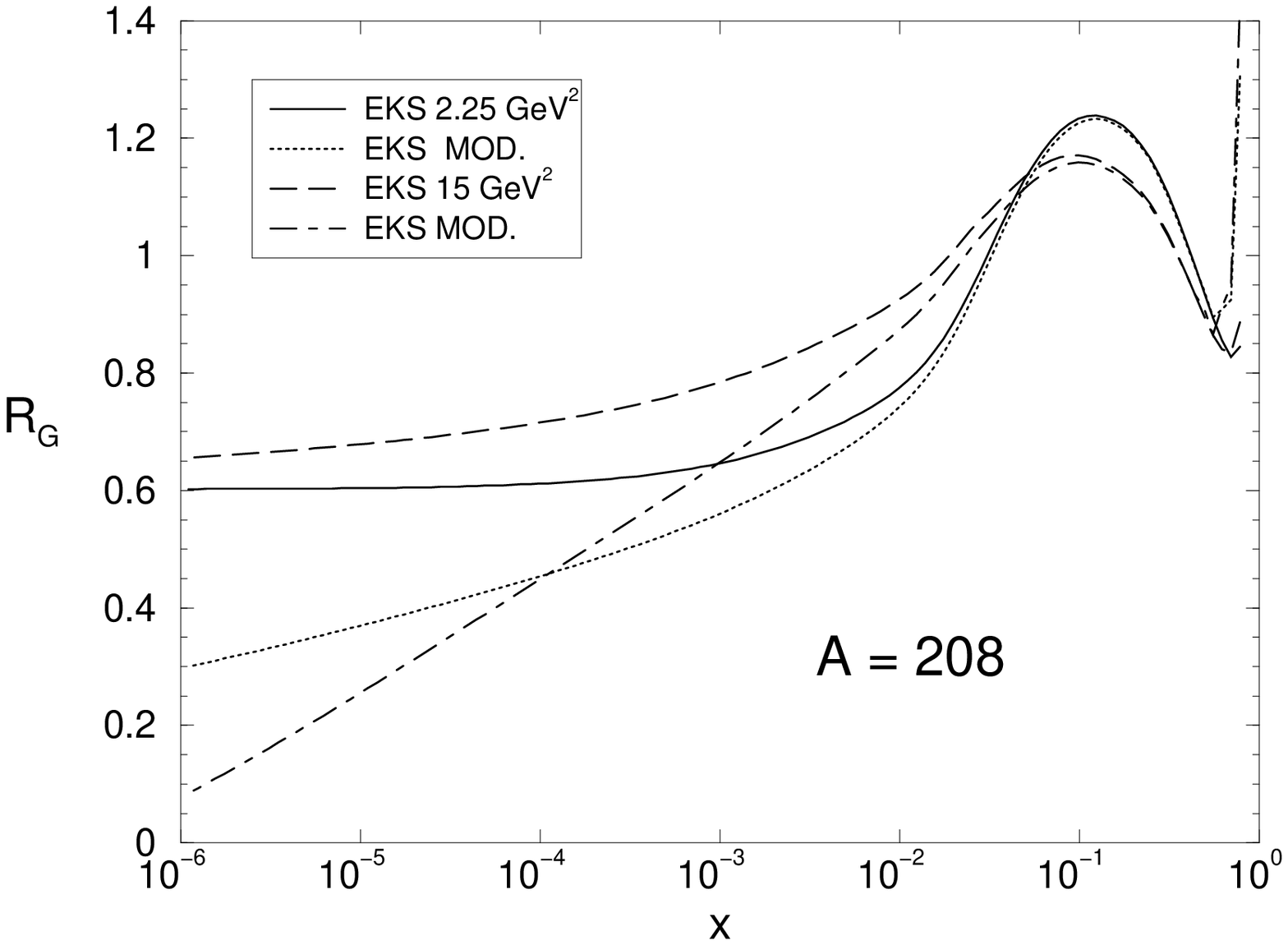,width=150mm}}
\caption{ }
\label{fig2}
\end{figure}

\vspace{1.0cm} Fig. \ref{fig2}: Comparison between the predictions of the
original EKS and EKS modified (denoted EKS MOD.   in  plot) for the ratio $%
R_G(x,Q^2)=xG_A/xG_N$ as function of the variable $x$ at different values
of $Q^2$. See text.

\newpage

%\begin{figure}
%\begin{tabular}{c c}
%\centerline{\psfig{file=gla2.eps,width=150mm}} 
%\end{tabular}
%\caption{First iteration of the Glauber-Mueller approach. The AGL equation 
%takes into account the interation of all partons  of the parton cascade  
%with several nucleons within of the nucleus.}
%\label{fig2}
%\end{figure}

\end{document}